\documentclass[prb,preprint,draft,showpacs,preprintnumbers,amsmath]{revtex4}
\usepackage{graphicx}
\preprint{CDW/2008}
\begin{document}
\DeclareGraphicsExtensions{.eps, .jpg}
\bibliographystyle{prsty}
\input epsf

\title{Observation of charge-density-wave excitations  in manganites} 
\author{A. Nucara$^{1}$, P. Maselli$^{1}$, P. Calvani$^{1}$, R. Sopracase$^{1}$,  M. Ortolani$^{2}$, G. Gruener$^{3}$, M. Cestelli Guidi$^{4}$, U. Schade$^{2}$, and  J. Garc\'ia$^{5}$}
\affiliation {$^1$CNR-INFM Coherentia and Dipartimento di Fisica, Universit\`a di Roma La Sapienza, Piazzale A. Moro 2, I-00185 Roma, Italy}
\affiliation{$^2$Berliner Elektronenspeicherring-Gesellshaft f\"ur Synchrotronstrahlung m.b.H., Albert-Einstein Strasse 15, D-12489 Berlin, Germany}
\affiliation {$^{3}$Laboratoire d'Electrodynamique des Mat\'eriaux Avanc\'es, UMR 6157 CNRS-CEA, 
Universit\'e Francois Rabelais, Parc de Grandmont, 37200 Tours, France}
\affiliation {$^{4}$Laboratori Nazionali INFN di Frascati, Via E. Fermi 40, Frascati, I-00044 Italy} 
\affiliation {$^{5}$Instituto de Ciencia de Materiales de Aragon and Departemento de Fisica de la Materia Condensada, Consejo Superior de Investigationes Cientificas y Universidad de Zaragoza, 50009 Zaragoza, Spain}

\date{\today}

\begin{abstract} 
In the optical conductivity  of  four different manganites with commensurate charge order (CO), strong peaks  appear in the meV range below  the ordering temperature $T_{CO}$. They are similar  to those reported for one-dimensional charge density waves (CDW) and are assigned to pinned phasons.  The peaks and their overtones allow one to obtain, for La$_{1-n/8}$Ca$_{n/8}$MnO$_3$ with $n$ = 5, 6, the electron-phonon coupling, the effective mass  of the CO system, and its contribution to the dielectric constant.  These results support a description  of the CO in La-Ca manganites in terms of moderately weak-coupling and of the CDW theory.
\end{abstract}


\maketitle

The  competition between the tendency of electrons to delocalize and the simultaneous presence of interactions like magnetism, Coulomb repulsion, and electron-phonon coupling is the origin of the extremely rich $x,T$ phase diagram of manganites like La$_{1-x}$Ca$_{x}$MnO$_3$ (LCMO) \cite{Cheong}. It also includes, for 0.5 $\le x < $ 0.85 and below an ordering temperature $T_{CO}$, charge-order (CO) phases  \cite{Dagotto}  which are usually associated with both orbital  ordering and antiferromagnetism (AF). For a long time these phenomena were described in terms of  stripes of localized charges,  where the mixed valence of manganese is split \cite{Good} between Mn$^{+3}$ ions surrounded by Jahn-Teller-distorted  O$^{-2}$ octahedra   and  Mn$^{+4}$ ions at undistorted sites.  This strong-coupling scenario has been recently questioned on the basis of experiments which indicate \cite{Subias}   charge disproportion  $<$1,   exclude strong localization \cite{Herrero}, and show \cite{Loudon1}  that the  charge modulation remains uniform when passing  from commensurate to incommensurate doping. A collective sliding of the charge system was also observed, at $x$ = 0.5, in not too strong electric fields \cite{Cox}. All these results  indicate that, in several manganites, the CO phase   can be described \cite{Milward} in terms of the Charge Density Waves (CDW) model \cite{Lee,Gruner},  which assumes weak electron-phonon coupling.
This hypothesis might be tested by measuring the  gap $2 \Delta$ which opens in the  real part of the optical conductivity $\sigma (\omega)$ below $T_{CO}$. Indeed, $2\Delta (T \simeq 0) / T_{CO}$ = 3.53 in the weak-coupling limit and increases  with the charge-lattice interaction. However,  different gap values are reported in the literature for similar compounds. They seem to  depend on the quality of the samples, on the extrapolation to zero of the mid-infrared $\sigma (\omega)$, on the addition of a pseudogap \cite{Calvani98,Noh02}. 

Besides the single-particle excitation at $\omega > 2 \Delta$, the CDW model predicts two collective modes \cite{Gruner} at $\omega << 2 \Delta$, the phason and the amplitudon. The former one is infrared active \cite{Gruner} and its  dispersion $\omega_p$ vs. $k$  is acoustic-like, with $\omega_p(\vec k = 0) =  \Omega_p$ = 0. However, either if the CDW is pinned to lattice impurities, or it  is commensurate with the lattice\cite{Gruner}, the phason  moves to  $\Omega_p > 0$.  The  amplitudon  is Raman active and its $\omega_a (\vec k)$  is similar to that of an optical phonon, with $\omega_a > \omega_p$ for any $\vec k$. 
Pairs of  such peaks  were  indeed detected at meV energies in one-dimensional conductors like the "blue bronze" family \cite{Degiorgi} K$_{0.3}$Mo$_{1-x}$W$_x$O$_3$. The lowest-energy member of the pair was assigned to the pinned phason, the other one to a  "collective bound resonance" due to the interaction between the collective mode itself and the impurities. The authors reported  for the CDW an effective mass $m^*/m_b$ varying from 700 to 800, where $m_b$ is the one-electron band mass.

In the manganite family one has the opportunity to study the excitation spectrum of  a multi-dimensional CDW. However, to our knowledge, experimental results have been reported up to now for Pr$_{0.7}$Ca$_{0.3}$MnO$_3$ only. By using  time-domain THz spectroscopy \cite{Kida}, both a  peak at 25 cm$^{-1}$ and a broad shoulder at lower frequencies were detected,  and attributed  to CDW collective excitations. The fact that both features survived at temperatures much higher that $T_{CO}$ was justified in terms of charge-order fluctuations on a local scale  \cite{Kida}. 

In the present paper we study, down to frequencies $\omega_{min}$ which range from  4.5 to 10 cm$^{-1}$, the optical  conductivity of  four manganites. One of them is a Nd$_{1/2}$Sr$_{1/2}$MnO$_3$ (NSMO) single crystal,  grown and  characterized as described in Ref. \onlinecite{Herrero},  and here measured in the $ab$ plane. The other ones are polycrystalline pellets of La$_{1-n/8}$Ca$_{n/8}$MnO$_3$ with $n$ = 5,6,7, namely with doping values where large single crystals are difficult to grow \cite{Noh02}.  They were obtained by  an organic gel-assisted citrate process \cite{Douy}, and controlled by X-ray powder diffraction. A further check on the sample with $n$ = 5  showed the expected change of slope in the  resistivity at $T_{CO}$ = 270 K. Both NSMO and  the LCMO's  with  $n$ = 5,6 are pseudocubic and, below $T_{CO}$, in a  highly symmetric AF phase  (CE) \cite{Pissas}. Therein,  $\sigma (\omega)$ was found to be independent of the crystal direction below 1 eV \cite{Tobe}.  Therefore, the  $ab$-plane conductivity of  NSMO can be correctly compared with that of the randomly oriented LCMO crystallites. In the case of $n$ = 7 the AF phase associated with the CO is uniaxial. However the structure is monoclinic and,  even if a single crystal were available, the response of different axes  could hardly be separated. As shown in Table I the four compounds have much different  $T_{CO}$, in order to establish a sound relation between the observed spectral features and their CO phases.

\begin{table} 

\caption{\label{Table I}
Transition temperatures $T_{CO}$ (from the literature) and charge-order parameters  of four manganites at 10 K  (from the present experiment). The symbols are explained in the text.}   
\begin{ruledtabular}

\begin{tabular}{cccccccc}
 
Sample   & $T_{CO}$ (K)   & $\Omega_p$ (cm$^{-1}$) & $\Omega_a$ (cm$^{-1}$) & $\lambda$ & $\epsilon^{CDW}_{0}$ & $\epsilon^{CDW}_{0,calc}$ & $m^*/m_b$  \\ 
\colrule
Nd$_{1/2}$Sr$_{1/2}$MnO$_3$ & 150 \cite{Popov} &  5.5 &	&   &   &  &   \\
La$_{3/8}$Ca$_{5/8}$MnO$_3$ & 270 \cite{Pissas}     &  10.5 - 14.5 & 40  &  0.7 & 100 & 90 & 400 \\
La$_{1/4}$Ca$_{3/4}$MnO$_3$  & 230 \cite{Pissas}  & 7.5 &  30 & 0.8 &  150 & 120 & 700 \\
La$_{1/8}$Ca$_{7/8}$MnO$_3$ & 130-150 \cite{Pissas,Li} 	  & 8.5	 &   &   &  &  &    \\
\end{tabular}
\end{ruledtabular}
\end{table}

\begin{figure}[tbp]
    \epsfxsize=8cm \epsfbox {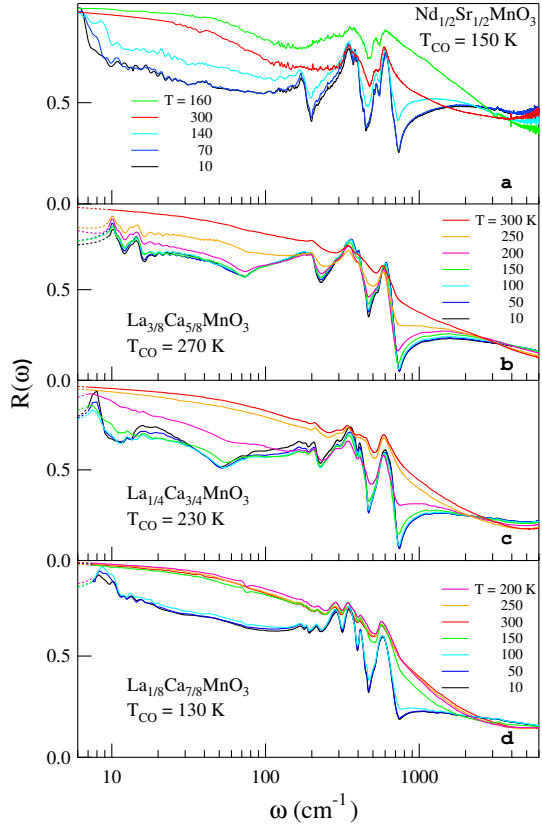}
\caption{Color online. Reflectivity at different temperatures of the single crystal of Nd$_{1/2}$Sr$_{1/2}$MnO$_3$  and of the polycrystalline pellets of La$_{1-n/8}$Ca$_{n/8}$MnO$_3$. The dashed lines show the Drude-Lorentz extrapolations of $R(\omega)$ from $\omega_{min}$ to $\omega$ = 0.}
\label{R}
\end{figure}

The reflectivity $R(\omega)$ of all samples was measured  at nearly normal incidence, after accurate polishing with sub-micron-thick powders. The sample thickness (2 mm) was such that   interference fringes due to multiple reflections were  negligible. However, their presencve both in the single crystal   (Fig.\ \ref{R}-a) and in the sample of Fig.\ \ref{R}-b, shows that the pellets are optically homogeneous at long wavelengths. The conventional radiation sources used between 20 and 6000 cm$^{-1}$ were replaced by the Coherent Synchrotron Radiation (CSR) of the BESSY storage ring \cite{Abo} from  $\omega_{min}$ to 30 cm$^{-1}$,  by just rotating the entrance mirror of the interferometer.  In the meV range  $R(\omega)$ was thus  measured\cite{diamond} with a 1\% error at a resolution of 0.5 cm$^{-1}$.  The samples were thermalized in a He-flow cryostat within $\pm$ 2 K, between 300 and 10 K. The  reference was the sample itself, coated with a gold  film evaporated \textit{in situ}. $R(\omega)$ was extrapolated  to $\omega = 0$ by accurate Drude-Lorentz fits  and  $\sigma(\omega)$  was obtained by standard Kramers-Kronig transformations.

The reflectivity of the four samples  is shown in Fig.\ \ref{R}. All spectra show  the transition to an insulating state below  $T_{CO}$, with the opening of an infrared  gap and the appearance of  strong phonon bands. The $R(\omega)$ of  NSMO in Fig.\ \ref{R}-a is intermediate between that reported for a polished crystal and that of  a cleaved crystal \cite{Takenaka02}. Below $T_{CO}$, in all samples of Fig. \ \ref{R} one also observes  novel spectral features in the meV range.  The lowest-frequency feature  is a sharp increase in $R$, by  10\% to 20 \%, around 10 cm$^{-1}$. In NSMO (Fig.\ \ref{R}-a),  it might indicate a very narrow free-carrier absorption. However, this  was ruled out after considering both  that the dc conductivity \cite{Takenaka02} of Nd$_{1/2}$Sr$_{1/2}$MnO$_3$ is vanishingly small at  low $T$,  and that one cannot fit its $R(\omega)$ to a Drude term at low $\omega$.   A comparison with the peaks  of the other three samples in Fig.\ \ref{R} suggests that also NSMO has a peak at  finite frequency, whose low-energy side is not seen because it is beyond $\omega_{min}$.  
In the La-Ca manganites a broad absorption is also detected below 100 cm$^{-1}$, with the same $T$-dependence as the main peak. At  $T_{CO}$ all the low-$\omega$ features disappear.This is not due to a shielding effect of the free carriers which, around $T_{CO}$, neither screen the phonons nor their weak shoulders.

\begin{figure}[tbp]
    \epsfxsize=8cm \epsfbox {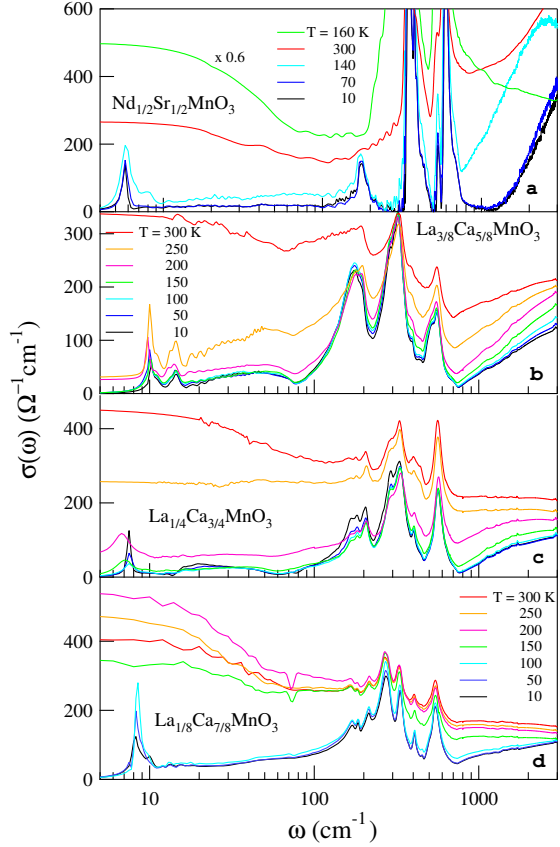}
\caption{Color online. Optical conductivity at different temperatures of the single crystal of Nd$_{1/2}$Sr$_{1/2}$MnO$_3$ and of the polycrystalline pellets of La$_{1-n/8}$Ca$_{n/8}$MnO$_3$.} 
\label{sigma}
\end{figure}

The optical conductivity of the four samples, shown in Fig.\ \ref{sigma},  exhibits a weak Drude term above $T_{CO}$.  In NSMO, which at $T > T_{CO}$ is ferromagnetic metallic,  $\sigma(\omega)$  increases upon heating, to decrease again above the Curie temperature $T_c \simeq$ 225 K (Fig.\ \ref{sigma}-a).  Below $T_{CO}$ an optical gap $2 \Delta$ opens in the $\sigma(\omega)$ of all samples. By smooth extrapolations  of the lowest-$T$ curves to $\sigma = 0$, one finds  $2 \Delta (T \simeq 0) \sim$  800 cm$^{-1}$ or 0.1 eV in the polycrystalline samples of LCMO, $\sim$ 0.2 eV in NSMO.  Correspondingly,  sharp conductivity peaks appear at the lowest frequencies   in all  panels of Fig.\ \ref{sigma}. Single peaks are observed in Nd$_{1/2}$Sr$_{1/2}$MnO$_3$ (Fig.\ \ref{sigma}-a), where the CO  is homogeneous with periodicity $\Lambda = 2a$ ($a$ = lattice constant), in La$_{1/4}$Ca$_{3/4}$MnO$_3$  (c) where  $\Lambda = 4a$  \cite{Cheong},  and in  La$_{1/8}$Ca$_{7/8}$MnO$_3$ (d) where $\Lambda$ is not known. Two well resolved peaks appear instead in La$_{3/8}$Ca$_{5/8}$MnO$_3$ (Fig.\ \ref{sigma}-b), where electron diffraction shows a coexistence of  $\Lambda = 2a$  (25\%) and  $\Lambda = 3a$ (75\%) \cite{Cheong}. The peak width increases with $T$ in a) and c), as expected, but neither in b) nor in d). These discrepancies may be attributed to the uncertainty in the extrapolation of  $R(\omega)$  to $\omega$ = 0 and to the changes with $T$ of the background reflectivity. We assign the peaks in the meV range to phasons in a  "pinned state", consistently with the presence in our samples of both impurities and commensurability. Their frequencies $\Omega_p$ ( see Table I) are all larger than  in \cite{Degiorgi}  K$_{0.3}$MoO$_3$ (3.3 cm$^{-1}$) and, in the three  samples with CE-type charge order, are found to scale with $T_{CO}$.

\begin{figure}[tbp]
    \epsfxsize=8cm \epsfbox {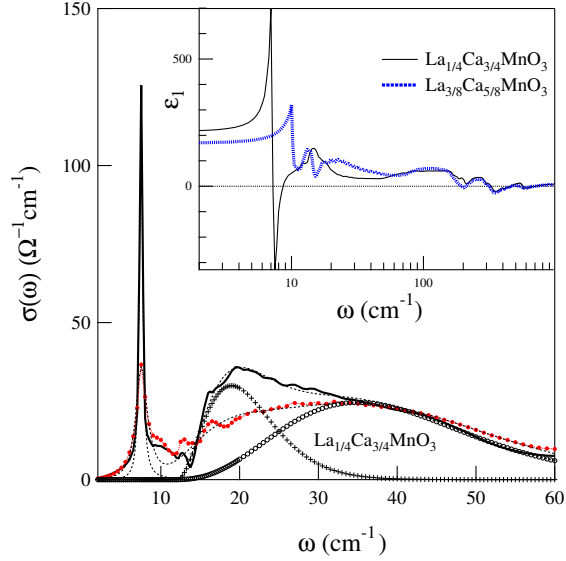}
\caption{Color online. Low-energy conductivity of La$_{1/4}$Ca$_{3/4}$MnO$_3$ at 10 K (solid line) and 100 K (dots) with the corresponding best fits  (thin-dotted lines). The side band is the sum of the overtone 2$\Omega_p$  and of the combination band $\Omega_p + \Omega_a$, as shown at 10 K by crosses and open circles, respectively. The inset shows the CDW contribution to the real part of the dielectric function $\epsilon_1 (\omega)$ in two samples at 10 K.}
\label{band}
\end{figure}

The broad band observed in the La-Ca compounds  at $\omega > \Omega_p$  approximately follows the $T$-dependence of the phason peak and disappears into the Drude continuum above $T_{CO}$.  In   La$_{1/4}$Ca$_{3/4}$MnO$_3$ (Fig.\ \ref{band}) it is shown to be the sum of two contributions. One of them broadens with $T$ like the phason peak at 7.5 cm$^{-1}$, while the other one is nearly $T$-independent. A similar fit was obtained for La$_{3/8}$Ca$_{5/8}$MnO$_3$. We assign those two features to the overtone $2\omega_p$ and to a phason-amplitudon combination band  $\omega_p + \omega_a$, respectively, basing on the following arguments: i) the CDW theory, once developed to the second order, implies a phase-amplitude 
mixing \cite{Lee}; ii)  a combination phason + amplitudon  is expected to be infrared active (as its irreducible representation should include that  of the phason); iii) the $T$-dependent peak shows at 10 K a sharp edge at $\simeq$  12 cm$^{-1}$, a value consistent with the minimum overtone frequency, 2$\Omega_p - \delta_{2p}$, for a reasonable anharmonic shift $\delta_{2p} \simeq$ 3  cm$^{-1}$; iv) the frequency predicted for the amplitudon at moderately weak coupling is $\sim$ 5 meV \cite{Gruner} or $\sim$ 40 cm$^{-1}$; v) the amplitudon should be more robust vs. the loss of coherence with temperature than the phason;  vi) both the overtone and the combination band are expected to be broad, as  their frequencies  result from any combination of  wavevectors $\vec k_p$ and $\vec k'_{a(p)}$, provided that $\vec k_p + \vec k'_{a(p)}  \simeq 0$. Similar bands are observed in the vibrational overtones of molecular crystals \cite{Calvani91}.

Both the pinned-phason  mode and the amplitudon have a nearly flat dispersion,  then a high density of states, close to $\vec k_a$ = 0.  If therefore   $\omega_p + \omega_a$ is peaked at $\Omega_p + \Omega_a$,  and  $\delta_{p+a} \simeq \delta_{2p}$, one obtains the values of $\Omega_a$  reported in Table I for $x$ = 5/8 and 3/4 (where the side band is better resolved). They  allow us to obtain a first description of the CO phase in manganites, including  the electron-phonon interaction strength $\lambda$. For $\vec k$ = 0 one has \cite{Gruner} 

\begin{equation}
\Omega_a^2 =  \lambda \omega^2(2k_F) \,,
\label{Omega}
\end{equation}

\noindent
where  $\omega(2k_F)$ is the acoustic-phonon frequency at twice the Fermi wavevector.  Here, for the CDW at $x$ =  3/4 (5/8) with period \cite{Cheong}  $4a$ (mainly 3$a$), $2k_F = \pi/2a$ (2$\pi/3a$). Calculations of the acoustic branches \cite{Rini} in LaMnO$_3$ provide $\omega(2k_F)$ = 36 (48) cm$^{-1}$, which gives $\lambda \simeq$ 0.8 (0.7). The assumptions of the CDW model are thus justified \textit{a posteriori}, independently of the evaluation of $\Delta$ and of its large uncertainty. 
The effective mass of the CDW is instead very sensitive to the gap value. Using Eq.\ \ref{Omega}, it can be written \cite{Gruner} as

\begin{equation}
\frac{m^*}{m_b}  \simeq \biggl [ \frac{2 \Delta}{\Omega_a} \biggr ]^2 \,,
\label{mass}
\end{equation}

\noindent
With the present $2 \Delta \sim$ 0.1 eV or 800 cm$^{-1}$ one obtains $m^*/m_b \simeq$ 400 (700) for $x$ = 5/8 (3/4) at 10 K, to be compared with  $m^*/m_b \simeq$ 800  reported for the one-dimensional CDW of  K$_{0.3}$MoO$_3$ \cite{Degiorgi}.

Equation\ \ref{mass} can also be used to theoretically predict another parameter which can be directly measured:  the CDW contribution to the dielectric constant $\epsilon^{CDW}_{0}$.  As here $\omega_{min}$ is lower than any CDW absorption,  the experimental $\epsilon_{1} (\omega_{min})$  at low $T$ measures $\epsilon^{CDW}_{0}$, after one subtracts  the phonon contributions and the high-frequency term $\epsilon_{\infty}$.  From the inset of Fig.\  \ref{band} we thus obtain $\epsilon^{CDW}_{0}$ = 100 (150) at $x$ = 5/8 (3/4). These results  can be compared with Eq. 3.22 of Ref. \onlinecite{Gruner}, that we write as
\begin{equation}
\epsilon^{CDW}_{0,calc} = 1 + \epsilon_{0,comb} + \biggl (\frac {\Omega_{pl}}{2\Delta}\biggr )^2 \biggl [ \biggl ( \frac{\Omega_a}{\Omega_p}\biggr )^2 +  4 \biggr ]  \,.
\label{epsilon}
\end{equation}

\noindent
Therein,  we have assumed that all  carriers condense  into the CDW ground state at $T  << T_{CO} $ and we have added the contribution $\epsilon_{0,comb}$ of the combination band.    In Eq.\  \ref{epsilon} one has  $2\Delta \sim$ 800 cm$^{-1}$, a plasma frequency $\Omega_{pl}$ ($T$ = 300 K) = 1200  (1540)  cm$^{-1}$, $\epsilon_{0,comb}$ = 34 (36) for LCMO with $x$ = 5/8 (3/4). By also using  the  $\Omega_p$ and $\Omega_a$ values of Table I, one obtains $ \epsilon^{CDW}_{0,calc}$ = 90 (120) at $x$ = 5/8 (3/4) in  very good  agreement with the experimental $\epsilon^{CDW}_{0}$  reported above.

In conclusion, we have first explored in the meV range the optical conductivity  of four manganites with different  $T_{CO} $, by using a coherent synchrotron source. We have found sharp peaks  which disappear above $T_{CO}$ and are similar to those reported for one-dimensional CDW's. They have been assigned to pinned phasons, followed by broad combinations of  phasons and amplitudons. All parameters extracted from the spectra, like the electron-phonon coupling and the CO-phase contribution to the dielectric constant, are consistent with  a description of the charge order in  La-Ca manganites in terms of charge density waves, even at commensurate doping. 

We wish to thank X. Blasco for providing  the NSMO crystal.

\end{document}